\documentclass[aps,twocolumn, floatfix,superscriptaddress]{revtex4}
\usepackage{graphicx}
\usepackage{hyperref}
\usepackage{amsmath}
\usepackage{amsfonts}
\usepackage{mathtools,bm}
\usepackage[amssymb]{SIunits}
\usepackage{color}
\usepackage{mathrsfs}

\newcommand{\ket}[1]{|{#1}\rangle}

\begin{document}
\flushbottom
\title{Negative-Mass Effects in Spin-Orbit Coupled Bose-Einstein Condensates}

\author{David Colas}
\email{d.colas@uq.edu.au}
\affiliation{ARC Centre of Excellence in Future Low-Energy Electronics Technologies, School of Mathematics and Physics, University of Queensland, St Lucia, Queensland 4072, Australia}

\author{Fabrice P. Laussy}
\affiliation{Faculty of Science and Engineering, University of Wolverhampton, Wulfruna St, Wolverhampton WV1 1LY, United Kingdom}
\affiliation{Russian Quantum Center, Novaya 100, 143025 Skolkovo, Moscow Region, Russia}

\author{Matthew J. Davis}
\affiliation{ARC Centre of Excellence in Future Low-Energy Electronics Technologies, School of Mathematics and Physics, University of Queensland, St Lucia, Queensland 4072, Australia}

\begin{abstract}  
  Negative effective masses can be realised by engineering the
  dispersion relation of a variety of quantum systems.  A recent
  experiment with spin-orbit coupled Bose-Einstein condensates has
  shown that a negative effective mass can halt the free expansion of
  the condensate and lead to fringes in the density [M.~Khamehchi
  \textit{et al.}, Phys. Rev. Lett. \textbf{118}, 155301 (2017)].
  Here, we show that the underlying cause of these observations is the
  self-interference of the wave packet that arises when only one of
  the two effective mass parameters that characterise the dispersion
  of the system is negative.  We show that spin-orbit coupled
  Bose-Einstein condensates may access regimes where both mass
  parameters controlling the propagation and diffusion of the
  condensate are negative, which leads to the novel phenomenon of
  counter-propagating self-interfering packets.
\end{abstract}

\pacs{} \date{\today} \maketitle

The most straightforward definition of mass in classical physics is
expressed by Newton's second law.  The acceleration $\mathbf{a}$ of an
object is related to the net force $\mathbf{F}$ acting upon it, with
the mass $m$ being the proportionality constant:
$\textbf{F}=m\mathbf{a}$. In this context, a particle with a negative
mass would behave strangely by accelerating in the opposite direction
of an applied force. This cannot happen in free space, but the concept
of mass can be extended beyond this simple scenario. In solid-state
physics, an \textit{effective} mass was originally introduced to
describe the motion of electrons in the periodic potential induced by
crystal lattices~\cite{kittel_book04a}. The effective mass $m^\ast$ is
related to the curvature of the 
dispersion relation, and for many quasiparticles this is nonparabolic,
leading to a mass that depends on the wavevector.  
A negative $m^\ast$ can occur, e.g., for semiconductor
holes near the top of a valence band.

A more general definition of mass is required when we consider both
the propagation and diffusion of wave packets. The dispersion can be
expanded in a Taylor series around $k_0$ as
$E(k) \approx E_0 + \hbar^2 k_0 (k-k_0)/m_1(k_0) +
\hbar^2(k-k_0)^2/[2m_2(k_0)] + \ldots{}$,
and the coefficients of each order term relate to a new mass parameter
that has certain characteristic effects on the
dynamics~\cite{larson05a,egorov09a}.
We  define
\begin{eqnarray}
m_{1}(k)&=&\hbar^2 k \left( \partial_k E(k)\right)^{-1}\,,\\
m_2(k) &\equiv& m^*=\hbar^2 (\partial_k^2 E(k))^{-1}\,.
\end{eqnarray}
The parameter $m_1$ is related to the classical motion of the wave
packet via the group velocity $v_g = \hbar k / m_1$.  The parameter
$m_2$ determines the acceleration of the packet when an external force
is applied, as well as its rate of diffusion~\cite{colas16a}.
For a purely parabolic dispersion we would find that $m_1 = m_2$, but
in other systems $m_1$ and $m_2$ can have different signs, be zero
or even become infinite.

A number of experimental platforms in physics now allow dispersion
engineering.  For example, exciton-polaritons produced in
semiconductor microcavities~\cite{kavokin_book17a} have a
non-parabolic dispersion that can be controlled by detuning the cavity
and the exciton modes, leading to a variety of exotic
effects~\cite{savvidis00a,amo09a,tosi11a}.  Recent theoretical and
experimental studies have shown that polariton wave packets can
exhibit a counter-intuitive flow resulting from a divergence of the
effective mass~$m_2$, in the form of Self-Interfering Packets
(SIP)~\cite{colas16a}, backflow \cite{ballarini17a} and superluminal
X-waves \cite{gianfrate18}.  When the wave packet spreads over this
singularity of the mass, it straddles regions of positive and negative
effective mass, effectively bouncing the packet back onto itself and
producing self-interference.

Another system that allows for the control of the dispersion of wave
packets is an atomic Bose-Einstein condensate (BEC). Early experiments
demonstrated dispersion engineering by loading a condensate into a
weak optical lattice~\cite{eiermann03a,morsch06a,eiermann04a}. Bright
solitons were subsequently realised in repulsive atomic and polariton
BECs by loading them into a quasi-momentum state with negative $m_2$
to counterbalance the effect of repulsive
interactions~\cite{eiermann04a,sich12a}. More recently, artificial
spin-orbit interactions in two-component BECs have allowed the
engineering of more complex dispersions through the control of the
Raman laser setup~\cite{linYJ11a,zhang12b,hamner15a}.  Interestingly,
this allows the possibility of generating negative regions for both
$m_1$ and $m_2$, which is not straightforward to achieve in polariton
systems.  In recent work, Khamehchi~\emph{et al.}\ have shown how the
peculiar dispersion relation of an atomic spin-orbit coupled
Bose-Einstein condensate (SOCBEC) can indeed lead to unconventional
wave packet dynamics, interpreted as ``negative-mass hydrodynamics'',
and reported phenomena such as self-trapping, soliton trains, and
dynamical instabilities~\cite{khamehchi17a}.

In this Letter, we clarify the role of the two effective mass
parameters $m_1$ and $m_2$ in determining the condensate dynamics in
the SOCBEC platform.  In particular, we show that the experimental
observation of inhibited expansion by Khamehchi \emph{et
  al.}~\cite{khamehchi17a} arises from a negative $m_2$ parameter, and
leads to the linear SIP phenomenon predicted earlier for
exciton-polariton BEC~\cite{colas16a}.  In the experiment, the
nonlinearity of the condensate then causes the interference fringes
from the SIP to transform into  solitons.  We further show that a
negative $m_1$ parameter can also be achieved in SOCBECs, and would
lead to the more striking phenomenology of a wave packet moving in the
opposite direction of its momentum.  In particular, we investigate the
dynamics in a regime where both $m_1$ and $m_2$ are negative, which
leads to the condensate splitting into two counter-propagating SIPs.
This is within reach of the current experimental platforms by simply
tuning the Raman parameters. A clear understanding of the underlying
mechanics of the wave packet dynamics can be obtained by performing
its wavelet decomposition. Our work thus provides a comprehensive
understanding of the effect of negative masses in SOCBECs.

\begin{figure}[h!]
  \includegraphics[width=\linewidth]{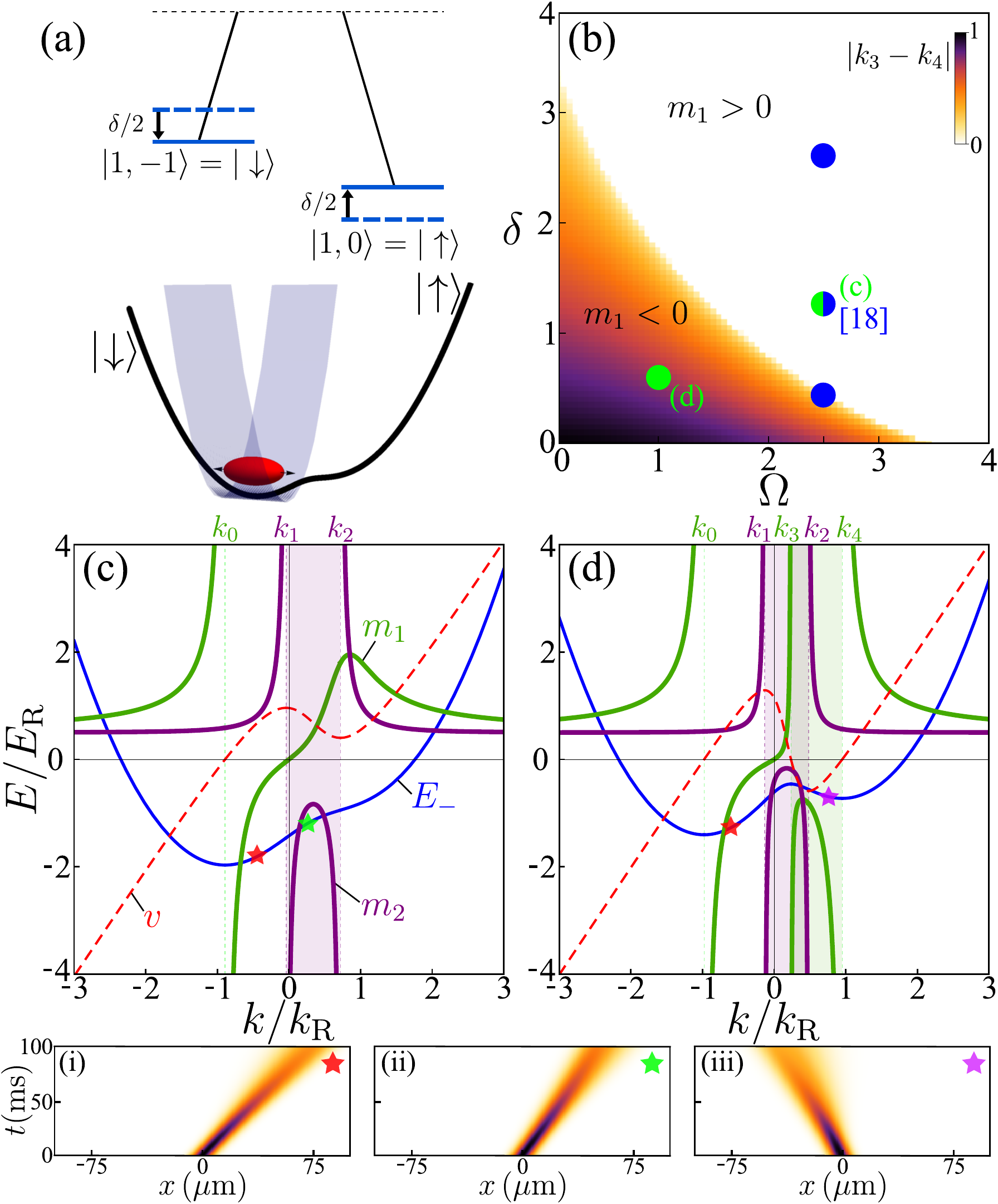}
  \caption{(Color online) (a) Schematic of the experimental
    configuration. The BEC initially resides at the bottom of the
    lower branch shaped by the spin-orbit coupling (see energy level
    diagram).  The trap is then released in the $x$ dimension to let
    the BEC expand. (b) Parameter space for $m_1$, defined by the momentum range $|k_3 - k_4|$ for which $m_1$ is
    negative. Blue dots: configurations considered
    in~\cite{khamehchi17a}. Green dots: configurations explored
    here. (c) SOCBEC dispersion properties for $\Omega=2.5$,
    $\delta=1.36$, $k_\mathrm{R}=1$ (from
    Ref.~\cite{khamehchi17a}). Blue line: lower branch $E_{-}$.  Green
    line: effective mass $m_1$.  Purple line: effective mass
    $m_2$. Red dashed line: group velocity $v$. The lower branch
    presents a region with $m_2<0$, where $v$ decreases with
    increasing $k$. (d) As for (c), but $\Omega=1$, $\delta=0.7$,
    $k_\mathrm{R}=1$, $E_{-}$ has regions with both $m_1<0$ and
    $m_2<0$, and $v$ has opposite sign to the momentum.  (i--iii)
    Examples of wave packet propagation, the coloured stars indicate
    where $E_{-}$ is excited.}
  \label{fig:1}
\end{figure}

We consider an identical spin-orbit coupling setup to Khamehchi
\emph{et al.}~\cite{khamehchi17a}, with a $^{87}\mathrm{Rb}$ BEC for
which two internal $5\mathrm{S}_{1/2}$ spin states are isolated,
leading to the so-called hyperfine pseudo-spin-up and pseudo-spin-down
states~\cite{linYJ11a}: $\ket{\uparrow}=\ket{F=1,m_F=0}$ and
$\ket{\downarrow}=\ket{F=1,m_F=-1}$.  Two Raman lasers are used to
couple the two states with a strength $\Omega$ and detuned by
$\delta/2$ from the Raman resonance, as shown in
Fig.~\ref{fig:1}(a). The system dynamics is described in units of
energy and momentum defined by the recoil energy
$E_\mathrm{R}=(\hbar k_\mathrm{R})^2/2m$ and Raman wavevector
$k_\mathrm{R}=2\pi/\sqrt{2}\lambda_\mathrm{R}$, where
$\lambda_\mathrm{R}$ is the Raman wavelength. For a homogeneous and
non-interacting gas, the system is described by the $k$-space
Hamiltonian~\cite{linYJ11a}
\begin{equation}
\label{eq:1}
H
= 
\begin{pmatrix}
 \frac{\hbar k (k + 2 k_\mathrm{R})}{2 m} +\frac{\delta}{2} & \frac{\Omega}{2}\\ 
 \frac{\Omega}{2} & \frac{\hbar k (k -2 k_\mathrm{R})}{2 m} -\frac{\delta}{2}
\end{pmatrix}\,,
\end{equation}
which acts on the spinor field
$\mathbf{\psi}=(\psi_\uparrow,\psi_\downarrow)^T$.  Diagonalising the Hamiltonian mixes the spin states and leads to upper ($+$) and lower ($-$) energy bands
\begin{equation}
\label{eq:2}
E_\pm (k) = \frac{\hbar k^2}{2m} \pm \sqrt{\left(\gamma k +\frac{\delta}{2}\right)^2 + \left(\frac{\Omega}{2}\right)^2}\, ,
\end{equation}
where $\gamma=\hbar k_\mathrm{R}/m $.
The dispersion relation for the \emph{lower} band for two sets of
Raman laser parameters $k_R$, $\Omega$ and $\delta$, are shown as a
blue line in Fig.~\ref{fig:1}(c, d). As both are clearly
non-parabolic, it is important to consider both the first and second
mass parameters, rather than only $m_2$ that was discussed in
Ref.~\cite{khamehchi17a}. The parameter space for $m_1$ is plotted in
Fig.~\ref{fig:1}(b), where we have identified the sets of parameters
considered in Ref.~\cite{khamehchi17a} as well as those of the
present Letter.  We emphasize that the dispersion relation in
Fig.~\ref{fig:1}(c) is identical to one set of SOC parameters
considered in Ref.~\cite{khamehchi17a}.

We now focus on the properties of the lower branch. An inflection
point of this branch corresponds to a change of sign of $m_2$, which
becomes infinite at the points $k_{1,2}$ meaning that wave packets
with this quasi-momentum do not diffuse~\cite{SuppMat}.
Similarly, one can define the points $k_{0,3,4}$ at which the
mass $m_1$ diverges~\cite{SuppMat}.
Most of the dynamics can now be understood regarding only the
$k$-dependent group velocity $v(k)$, directly shaped by $m_1$ and
$m_2$~\cite{SuppMat}.
As can be seen in Fig.~\ref{fig:1}(c,d), the linear parts of the
dispersion, when $m_2$ diverges at $k_{1,2}$, lead to a local minimum
or maximum for the group velocity. Similarly, $v$ is zero when the
dispersion is locally flat at the points $k_{0,3,4}$, and takes
negative values between $k_{3}$ and $k_4$, where $m_1<0$. A negative
$m_1$ here corresponds to the packet moving in the opposite direction
to the applied impulse, thus reversing the sign of the group
velocity. We show in Fig.~\ref{fig:1}(i-iii) typical examples of wave
packet propagation when exciting the branch at different
quasi-momenta. 

For the dispersion shown in Fig.~\ref{fig:1}(c), consider applying an impulse to move from the red (i) to the green (ii) star, where $m_2<0$.  The wave packet decelerates, but keeps propagating in the same direction.
Conversely, for the dispersion shown in Fig.~\ref{fig:1}(d), applying an impulse to move from the red (i) to the purple star (iii) (where $v<0$), one sees that the wave packet not only slows down, but actually reverses direction.
We have provided an online interactive plot of the
dispersion as a function of its key parameters $\delta$, $\Omega$ and
$k_{\mathrm{R}}$~\cite{SuppMat2}.

We now analyse one-dimensional simulations of the dynamics of SOCBEC
expansion in the context of $m_1$ and $m_2$ using the experimental
parameters for $\delta=1.36$ and $\Omega=2.5$ from
Ref.~\cite{khamehchi17a,footnote1}. The dynamics of a 1D BEC initially
positioned at the bottom of the lower branch and released from a
harmonic trap in one direction can be described by a single-band
Gross-Pitaevskii equation:
\begin{equation}
\label{eq:3}
i\partial_t\psi(x)= \mathscr{F}^{-1}_x[E_{-}(k)\psi(k)] +g|\psi(x)|^2\psi(x)\,,
\end{equation}
where $E_{-}(k)$ is the lower branch dispersion derived in
Eq.~\ref{eq:2} and shown in Fig.~\ref{fig:1}(c), $\mathscr{F}^{-1}_x$
is the inverse Fourier transform, and $g$ the effective 1D interaction
strength.  

We initially focus on the linear dynamics by setting
$g=0$ so that we can study the free wave
packet propagation with this dispersion, for which only $m_2$ is negative.
\begin{figure}[t!]
  \includegraphics[width=\linewidth]{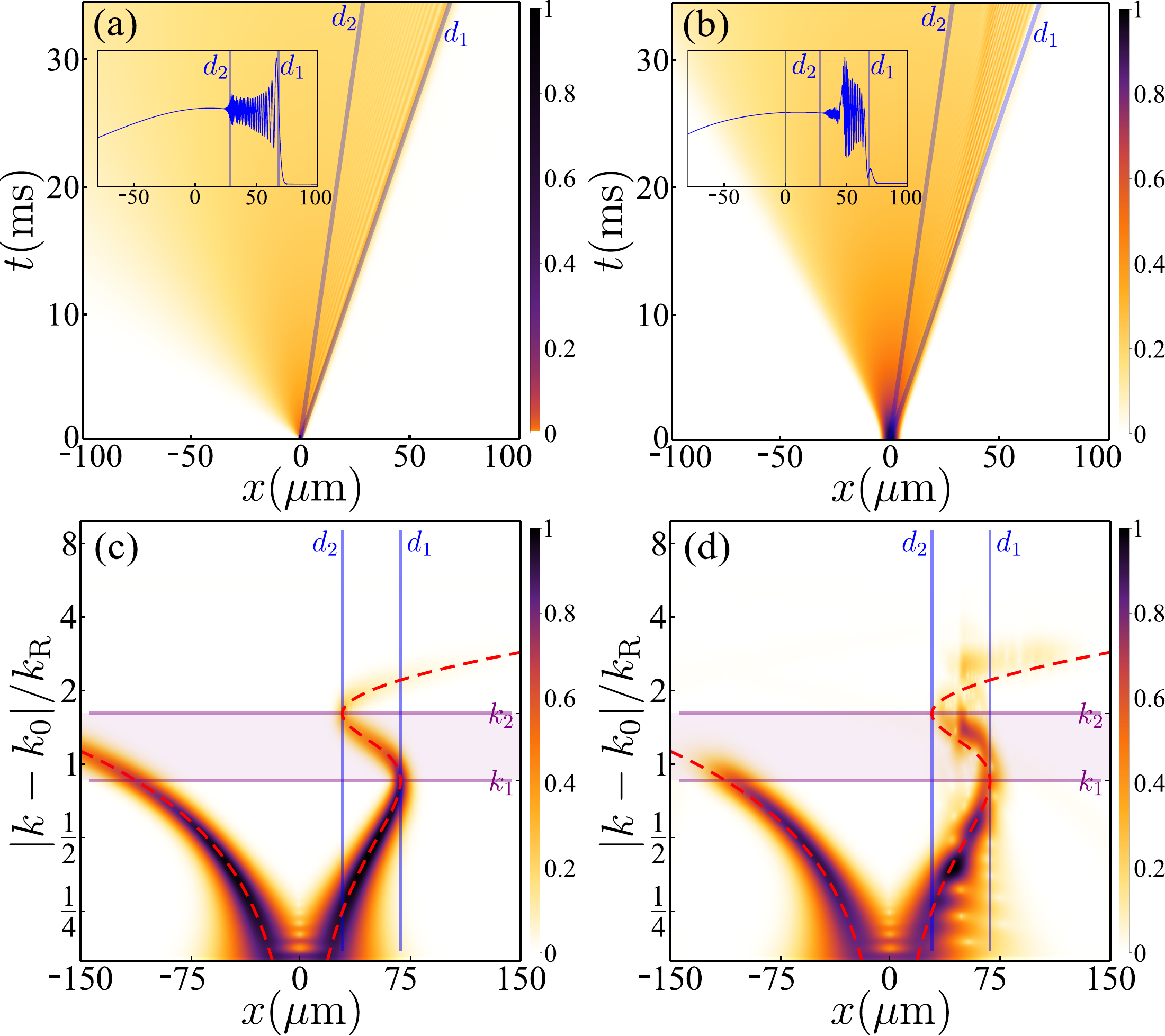}
  \caption{(Color online) BEC expansion with the SOC dispersion of
    Fig.~\ref{fig:1}(c). The first column shows the linear case with
    $g=0$, and the second column the interacting case. (a--b) Spacetime evolution of $|\psi(x,t)|^2$, with a condensate initial size of (a) $\unit{0.25}{\micro\meter}$,
    and (b) $\unit{5}{\micro\meter}$ Thomas-Fermi radius, obtained from the ground state of a $10^4$ particles BEC released from a $\unit{100}{\hertz}$ trap. Insets: Density at $t=35$ ms. (c--d) Wavelet
    decomposition of $\psi(x)$ at $t=\unit{35}{\milli\second}$. Vertical blue lines: limits of the diffusion cone in
    $x$. Purple horizontal lines: position of the
    inflection points in $k$. Red dashed curve: classical displacement $d(k)$ of each $k$-wave vector. An animation of this figure is provided as Supplementary Video S1~\cite{SuppMat}.}
  \label{fig:2}
\end{figure}
We begin with a narrow Gaussian wave packet
($\sigma_x=\unit{0.25}{\micro\meter}$) centered at the minimum of the
branch, so that its momentum spread encompasses the range $k_1$--$k_2$
~\cite{footnote2}. The spacetime evolution of the wave function
$|\psi(x,t)|^2$ is plotted in Fig.~\ref{fig:2}(a) and shows a distinct
interference pattern spatially confined in the diffusion cone defined
by $d_{1,2}(t)=v(k=k_{1,2})t$.
An enlightening method to visualise the self-interference effect is to
plot the wave function density in the $x$-$k$ plane by performing the
Wavelet Transform
(WT)~$\mathbb{W}(x,k)=(1/\sqrt{|k|})\int_{-\infty}^{+\infty}\psi(x)\mathcal{G}^\ast
[(x-x_0)/k]\mathrm{d}x$~\cite{debnath_book15a}.
Recent studies have shown that the WT can be applied to analyse
complex interacting wave packets
dynamics~\cite{baker12a,colas16a}. Unlike the usual Fourier transform
based on the decomposition of the signal into a sum of delocalised
functions (sine and cosine), the WT uses localised wavelets. Here we
choose the Gabor wavelet family, with Gaussian-like
functions~$\mathcal{G}^*$, and a high central frequency ensuring good
resolution near the inflection points in the
dispersion~\cite{mark97a}.
Further details regarding the WT for wave packets are described in
the Supplementary Material~\cite{SuppMat}.

 Figure~\ref{fig:2}(c) shows
the wavelet energy density $|\mathbb{W}(x,k)|^2$ 
at $t=\unit{35}{\milli\second}$. The inflection point
momenta $k_{1,2}$ are indicated in $k$ (purple lines), and the
boundaries $d_{1,2}$ of the diffusion cone in $x$ (blue lines). We  also
plot the displacement $d(k)=v(k)t$ associated with each
$k$-wave vector (red dashed line). From this simple picture, one can now
understand the origin of the self-interference effect: different wave
vectors of the packet travel with the same velocity, hence overlapping
in real space and interfering. This happens only when the wave packet
spreads over an inflection point of the branch. This phenomenology is a universal consequence of the shape of the dispersion, and can therefore be equally
encountered for exciton-polariton and atomic condensates, despite
their operating time scales that differ by nine orders of magnitude.

Practically, it is challenging to form a SOCBEC with such a broad
spread in momentum. One way to overcome this difficulty is to load the
packet directly in the inflection points region by imparting it with
the appropriate momentum.
A second way, used in the experiment of Khamehchi \textit{et
  al.}~\cite{khamehchi17a}, is to release the BEC from the trap
leading to a broadening of the wave packet in $k$-space due to the
conversion of interaction energy to kinetic energy. This is an
effective way to push some components of the wave packet into the
negative $m_2$ region.  Further details of this approach are presented
in the Supplementary Material~\cite{SuppMat}.

Here we simulate the expansion dynamics of an interacting system
initially in a $\unit{100}{\hertz}$ harmonic trap with a Thomas-Fermi
radius of 5 $\mu$m corresponding to approximately $10^4$ atoms.  This
is more tightly confined with fewer atoms than the experiment of
Ref.~\cite{khamehchi17a}, which had a trap frequency of
$\unit{26}{\hertz}$, $10^5$ atoms, and a Thomas-Fermi radius of $23$
$\mu$m, but enables a direct comparison with the $g=0$ case.
The results for the spacetime density and WT are shown in
Fig.~\ref{fig:2}(b,d).  Here the interference only becomes visible
after a finite time, which is that needed for the expansion of the
wave packet to reach $k_1$. The presence of self-interference is again
confirmed by the WT analysis in Fig.~\ref{fig:2}(d).  Due to the
interactions, the self-interference pattern differs slightly from the
non-interacting case, and the energy density distribution gets spread
around $d(k)$, since this curve accounts for the packet's
$k$-components displacement from $t=0$. The initial packet is here 20
times larger and only the low $k$-region is significantly populated at
$t=0$. We provide analysis of the exact experimental situation of
Khamehchi \textit{et al.}~\cite{khamehchi17a} in the Supplementary
Material~\cite{SuppMat}. Their configuration has a larger nonlinearity
which results in the population of $k$-components above $k_2$, which
travel with a velocity larger than $v(k_1)$.  This results in the
density ``leaking out'' of the diffusion cone ($d_1$), and was
previously identified as being due to dynamical
instability~\cite{khamehchi17a}. The limited diffusion of the
condensate as shown in Fig.~\ref{fig:2}(b), which might appear as a
``self-trapping'' effect, can be overcome if high enough momenta
($k>k_2$) are reached, as is the case in Ref.~\cite{khamehchi17a}.
\begin{figure}[t!]
  \includegraphics[width=\linewidth]{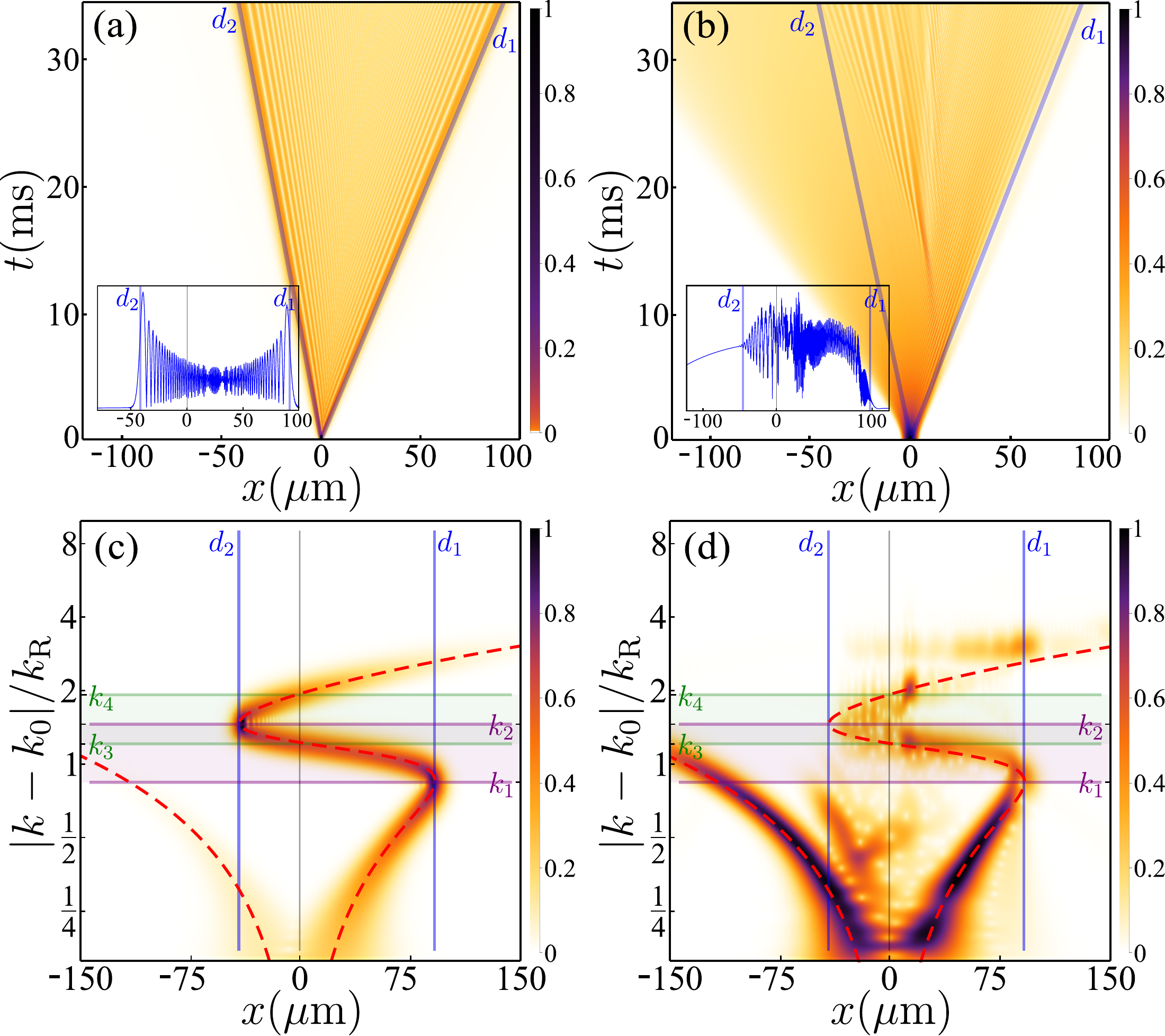}
  \caption{(Color online) BEC expansion with the SOC dispersion of
    Fig.~\ref{fig:1}(d) with both $m_1<0$ and $m_2<0$. The first
    column shows the noninteracting case with $g=0$, and the second
    column the interacting case. (a--b) Spacetime evolution of
    $|\psi(x,t)|^2$, with a condensate initial size of (a)
    $\unit{0.25}{\micro\meter}$, and (b) $\unit{5}{\micro\meter}$
    Thomas-Fermi radius, obtained from the ground state of a
    $\approx 2.5\times 10^4$ particles BEC released from a
    $\unit{100}{\hertz}$ trap. Insets: density at $t=35$ ms. (e--f) Wavelet decomposition of
    $\psi(x)$ at $t=\unit{35}{\milli\second}$. The color code is the
    same as Fig.~\ref{fig:2}(c,d) with additional horizontal green
    lines delimiting the $m_1<0$ region.  An animation of this figure
    is provided as Supplementary Video S2~\cite{SuppMat}.}
  \label{fig:3}
\end{figure}

We now study the SOCBEC dispersion in a more interesting configuration
that exhibits a stronger type of negative mass effect, shown in
Fig.~\ref{fig:1}(d).
This can be obtained by reducing the Raman coupling $\Omega$ and
detuning $\delta$.  In this case the mass $m_1$ is also negative in
the momentum range $k_3$--$k_4$ for which $v<0$.  Again, we first
consider the non-interacting case ($g=0$), but this time with a
Gaussian packet centered on $k=k_2$ with a width
$\sigma_x=\unit{0.25}{\micro\meter}$. As shown in Fig.~\ref{fig:3}(a),
the wave packet exhibits a double SIP effect during its propagation as
the initial state spreads over both $k_1$ and $k_2$. The most notable
feature is the position of the second SIP, whose diffusion is limited
by $d_2=v(k_2)t$ in the $x<0$ region, as $k_3<k_2<k_4$. The packet
is now composed of two sub-packets, each carrying and propagating a
SIP in opposite directions. This can be seen in the WT in
Fig.~\ref{fig:3}(c), where we have added the boundaries $k_{3,4}$
(green lines) delimiting the $m_1<0$ region. One can see how the
wavelet energy density in the momentum range $k_3$--$k_4$ is only displayed
in the $x<0$ region. This corresponds to the
packet's $k$-components experiencing backward propagation.

The double SIP behaviour can also be observed in the expansion of an
interacting BEC. We again begin with the condensate ground state in a
$\unit{100}{\hertz}$ trap with a Thomas-Fermi radius of
$\unit{6}{\micro\meter}$, corresponding to $\approx 2.5\times 10^4$
atoms, leading to the population of higher momenta as compared to the
previous case.  As before, the SIP effect is present within the
overall diffusing packet as can be seen in Fig.~\ref{fig:3}(b), and in
the WT shown in Fig.~\ref{fig:3}(d), one can see that the SIP is
``delayed'' as compared to the non-interacting case. Time-animated
videos of $|\psi(x,t)|^2$ with $|\mathbb{W}(x,k)|^2$ for both the
linear and interacting cases as presented in Fig.~\ref{fig:2}
and~\ref{fig:3} are provided in Ref.~\cite{SuppMat}. In experiments,
translating optical lattices or Bragg pulses could be used to impart a
momentum to the condensate in order to more clearly exhibit the SIP
effect~\cite{hamner15a}. Other nonlinear features like shock waves or
soliton trains observed by Khamehchi \textit{et
  al.}~\cite{khamehchi17a}, may appear \textit{a posteriori} as the
self-interference process induces large oscillations of the condensate
density, and thus provide a breeding ground for these excitations. In
this experiment, both these linear and nonlinear effects are intertwined and 
cannot be clearly separated in the dynamics.
The limited diffusion of the condensate, sometimes
described as a nonlinear self-trapping effect in the
literature~\cite{anker05a,alexander06a,xue08a}, can instead be viewed
in this case as a consequence of peculiar dispersion relations
containing inflection points and regions of negative effective mass.
Although the condensate diffusion is strongly affected, it is not
bounded and normal diffusion can still occur as the dispersion returns
to being parabolic at higher momenta.

In conclusion, we have shown that SOCBECs provide an excellent
platform to engineer dispersion relations, allowing the creation of
regions of negative effective mass for both parameters $m_1$ and $m_2$
that govern wave packet dynamics. The mass $m_1$ leads to a negative
group velocity for a positive impulse, while the mass $m_2$ leads to
self-interference in the wave packet.  Self-interference alone is a
direct consequence of the linear dynamics where dispersion relations
contain inflection points. The nonlinearity of the BEC can further add to
the phenomenology, in particular by populating such regions in the
momentum space, and allowing self-interference to subsequently lead to the
formation of  solitons.  
The wave packet dynamics
for which both mass parameters are negative is within reach of the
SOCBEC platforms. This would result in the formidable phenomenology of
moving an object in the direction opposite to which it was pushed.

\begin{acknowledgements}
  This research was supported by the Australian Research Council
  Centre of Excellence in Future Low-Energy Electronics Technologies
  (project number CE170100039) and funded by the Australian
  Government.  It was also supported by the Ministry of Science and
  Education of the Russian Federation through the Russian-Greek
  project RFMEFI61617X0085 and the Spanish MINECO under contract
  FIS2015-64951-R (CLAQUE).
\end{acknowledgements}
%


\pagebreak

\onecolumngrid
\begin{center}
  \textbf{\large Supplemental Material:\\
Negative-Mass Effects in Spin-Orbit Coupled Bose-Einstein Condensates}\\[.2cm]
  David Colas,$^{1,*}$ Fabrice P. Laussy,$^{2,3}$ and Matthew J. Davis$^1$\\[.1cm]
  {\itshape ${}^1$ARC Centre of Excellence in Future Low-Energy Electronics Technologies,\\ School of Mathematics and Physics, University of Queensland, St Lucia, Queensland 4072, Australia\\
  ${}^2$Faculty of Science and Engineering, University of Wolverhampton,\\Wulfruna St, Wolverhampton WV1 1LY, United Kingdom\\
  ${}^3$Russian Quantum Center, Novaya 100, 143025 Skolkovo, Moscow Region, Russia\\}
  ${}^*$Electronic address: d.colas@uq.edu.au\\
(Dated: \today)\\[1cm]
\end{center}
\twocolumngrid

\setcounter{equation}{0}
\setcounter{figure}{0}
\setcounter{table}{0}
\setcounter{page}{1}
\renewcommand{\theequation}{S\arabic{equation}}
\renewcommand{\thefigure}{S\arabic{figure}}
\renewcommand{\bibnumfmt}[1]{[S#1]}
\renewcommand{\citenumfont}[1]{S#1}

\section{Introduction}
In this Supplementary Material, we provide further details regarding
the analysis and the expansion dynamics of the SOCBEC in the context
of Self-Interfering Packets. In Section II we provide a detailed
analytical analysis of the properties of the SOC dispersion. In
Section III we describe the Wavelet Transform and its properties.
 In Section IV we present the energy transformations in the expansion of the
interacting condensate, and we show the quantitative differences
between the cases of parabolic and SOC dispersion. In Section V we directly apply our analysis to 1D simulations using identical parameters to the SOCBEC expansion dynamics presented by Khamehchi \textit{et al.}~[18].  Finally,  Section VI describes the three videos accompanying this Supplementary Material to more clearly show the complex dynamics occurring in the SOCBEC. The
Equations and Figures from the main text are here quoted with numbers
whereas those from the Supplementary are prefixed by ``S''.
\section{SOCBEC dispersion analysis}
The inflection points of the SOCBEC dispersion relation identify the wavevectors $k_{1,2}$ where the effective mass $m_2$  is infinite. Their location can be found  by solving $1/m_2 = \partial_k^2 E_{-}(k)=0$, giving the following result:
\begin{equation}
\label{eq:s_0}
k_{1,2}= \frac{\delta}{2 \gamma} \pm \sqrt{\left(  \frac{m \Omega^2}{4 \gamma \hbar}  \right)^{\frac{2}{3}} -\left( \frac{\Omega}{2 \gamma}  \right)^2}\,.
\end{equation}
This expression also provides a condition on the existence of the inflection points in the dispersion relation, as the term under the square root of Eq.~\ref{eq:s_0} has to be positive. Thus  the dispersion relation possesses inflection points if:
\begin{equation}
\label{eq:s_01}
\frac{\hbar^2 k_\mathrm{R}^2}{2 m} <\frac{\hbar \Omega}{4}\,.
\end{equation}
The points $k_{0,3,4}$ at which the effective mass $m_1$ diverges  can be found by solving $\partial_k E_{-}(k)= v(k)=0$. The analytical expression for the group velocity is:
\begin{equation}
\label{eq:s_02}
v=\frac{\hbar k}{m} - \gamma \bigg/ \sqrt{\left(\frac{\Omega}{2 k \gamma -\delta}\right)^2 +1}\,.
\end{equation}
One can see that for  $k \gg \Omega/\gamma$ the dispersion is parabolic again, and the velocity grows linearly with the momentum.  Analytical expressions exist for the points $k_{0,3,4}$ but they are too cumbersome to reproduce here. The point $k_0$ corresponds to the bottom of the SOCBEC dispersion and always exist. The effective mass $m_1$ is negative between the points $k_3$ and $k_4$ (when they exist), which corresponds to a negative velocity $v$. Figure~1(b) of the main text summarizes the parameter space for which $m_{1}$ is positive or negative.

\section{Wavelet Transform}
\begin{figure}[h!]
  \includegraphics[width=\linewidth]{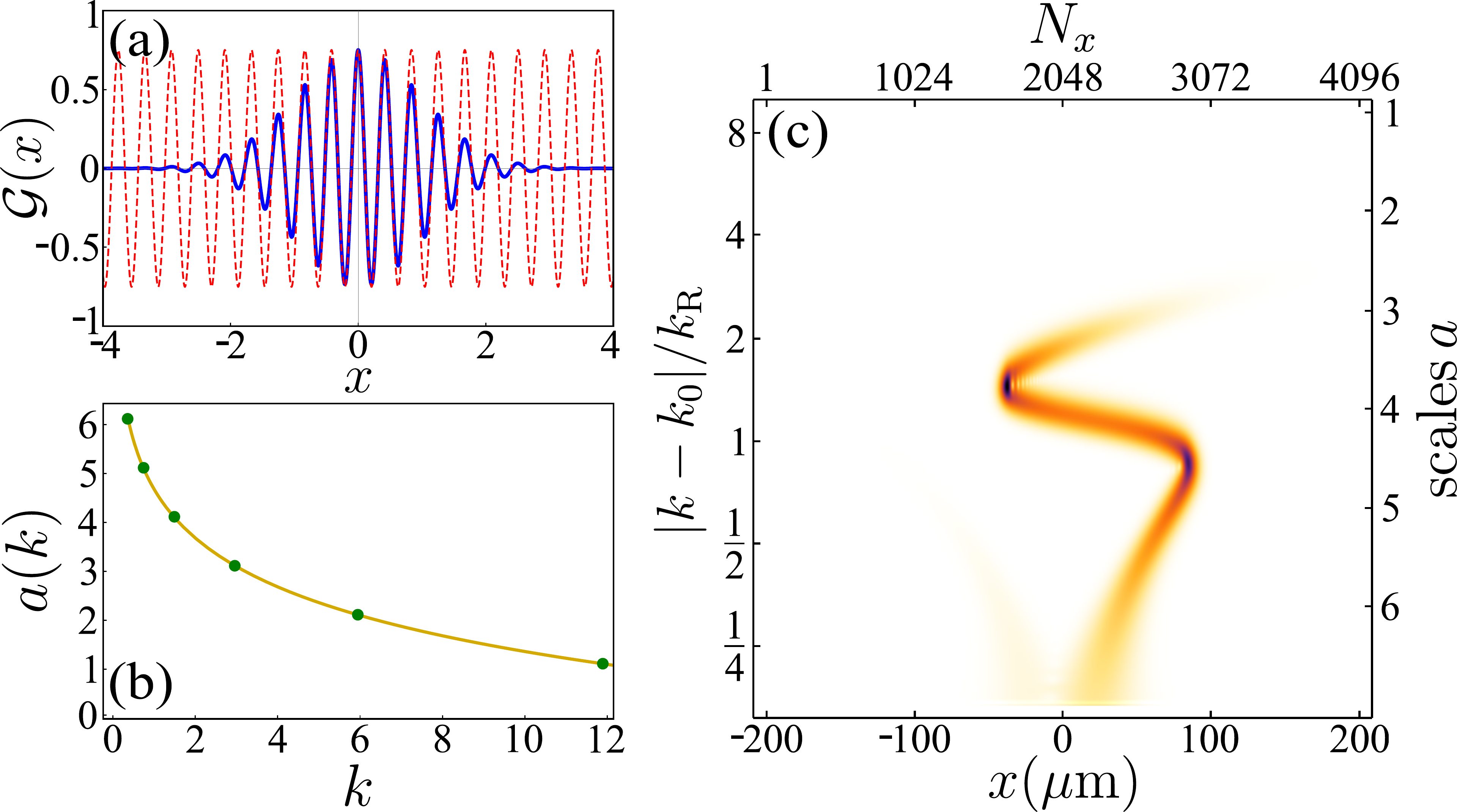}
  \caption{(a) Real part of the Gabor wavelet with a frequency $\omega=15$ (blue line) and its corresponding central frequency (dashed red line). (b) Log base 2 scaling  between scales $a$ and momenta $k$. (c) Example of a scalogram showing the correspondences between: (i) Scales and momentum units; (ii) Number of points $N_x$ and space units.}
  \label{fig:s_1}
\end{figure}
The Wavelet Transform (WT) is commonly used in signal processing to obtain a signal representation both in time and frequency. This transform can also be  applied to any complex wave function $\psi(x)$ in order to obtain its representation in both position $x$ and momentum $k$. 
The Wavelet Transform is:
\begin{equation}
\label{eq:s_1}
\mathbb{W}(a,b)=(1/\sqrt{a})\int_{-\infty}^{+\infty}\psi(x)\mathcal{G}^\ast [(x-b)/a]\mathrm{d}x\,,
\end{equation}
where $a$ is the scale and $b$ the translation parameter. In physical
terms, this quantity measures the cross-correlation between the wave
function $\psi(x)$ and a wavelet $\mathcal{G}(x)$ that is scaled by $a$ and
translated in space by $b$. This map
$|\mathbb{W}(x,k)|^2$, also called \textit{scalogram}, indicates the
cross-correlation between position and momentum of the wave function. In the
context of this paper, the WT is a remarkable tool to understand the
complex wave packet dynamics that occurs in SOC systems.  For this
study, we work with the Gabor Wavelet:
\begin{equation}
\label{eq:s_2}
\mathcal{G}(x)=\sqrt[4]{\pi}\exp(i \omega x)\exp(-x^2/2)\,,
\end{equation}
where $\omega$ is the wavelet's frequency. The choice of this wavelet
is natural since the Gabor wavelet---a Gaussian function with a
well-defined frequency---corresponds to the archetype of a propagating
quantum wave packet, embedding motion and diffusion in a transparent
way~[25]. 
The central frequency of the Gabor wavelet can be
well approximated by:
\begin{equation}
\label{eq:s_3}
f_c=\frac{2\pi(\omega +\sqrt{w^2 +2})x}{4\pi}\,.
\end{equation}
The wavelet oscillations thus match with the oscillatory function
$D=\pi^{-1/4}(\cos(f_c x)+i \sin(f_c x))$. In Fig.~\ref{fig:s_1}, the
real part of the Gabor wavelet is plotted in blue and the function
$\Re(D)$ in dashed red. In our numerical computations, we choose a
relatively high wavelet frequency $\omega=15$ in order to obtain high
cross-correlations in the inflection points region. We can now obtain the
correspondence between the scales $a$ of the WT and the signal
frequency (here the momentum) with:
\begin{equation}
\label{eq:s_4}
k_a=\frac{f_c}{a N_x}\,,
\end{equation}
where $N_x$ is the number of points in the spatial grid, corresponding
to the signal sampling rate. The scaling law that we use between
scales and momentum is shown in Fig.~\ref{fig:s_1}(b). It follows a
log base 2 scale, meaning that the momentum at the next scale is doubled
(like the octaves in music). Low scales thus correspond to high
momenta and vice versa. Here, each scale (octave) is also subdivided in 20 voices.
 An example of a scalogram showing the correspondence between
scales (momenta) and grid points (position) is shown in
Fig.~\ref{fig:s_1}(c).

\section{Energy conversions of the condensate}

The total energy functional for the Hamiltonian of Eq.~(7) can be
written as $E_{\mathrm{tot}}=E_{\mathrm{P}}+E_{\mathrm{NP}}+E_{\mathrm{int}}$, accounting for the kinetic
energy from the parabolic and non-parabolic parts of the dispersion
[see Eq.~(4)], and for the interaction energy due to the nonlinear
term of Eq.~(7), respectively. The first and the last terms recover
the total energy of the GPE in free space:
\begin{equation}
\label{eq:s_8}
E_{\mathrm{P}} + E_{\mathrm{int}}= \int\left(\frac{\hbar^2}{2m}|\nabla \psi(x)|^2 +\frac{g}{2}|\psi(x)|^4\right)\mathrm{d}x\,.
\end{equation}
The contribution from the non-parabolic part of the dispersion is most easily computed in momentum space:
\begin{equation}
\label{eq:s_9}
E_{\mathrm{NP}}= \int -\frac{1}{2}\sqrt{\Omega^2(k -2 k_\mathrm{R})^2 + \delta^2}|\psi(k)|^2\mathrm{d}k\,.
\end{equation}
\begin{figure}[t!]
  \includegraphics[width=0.8\linewidth]{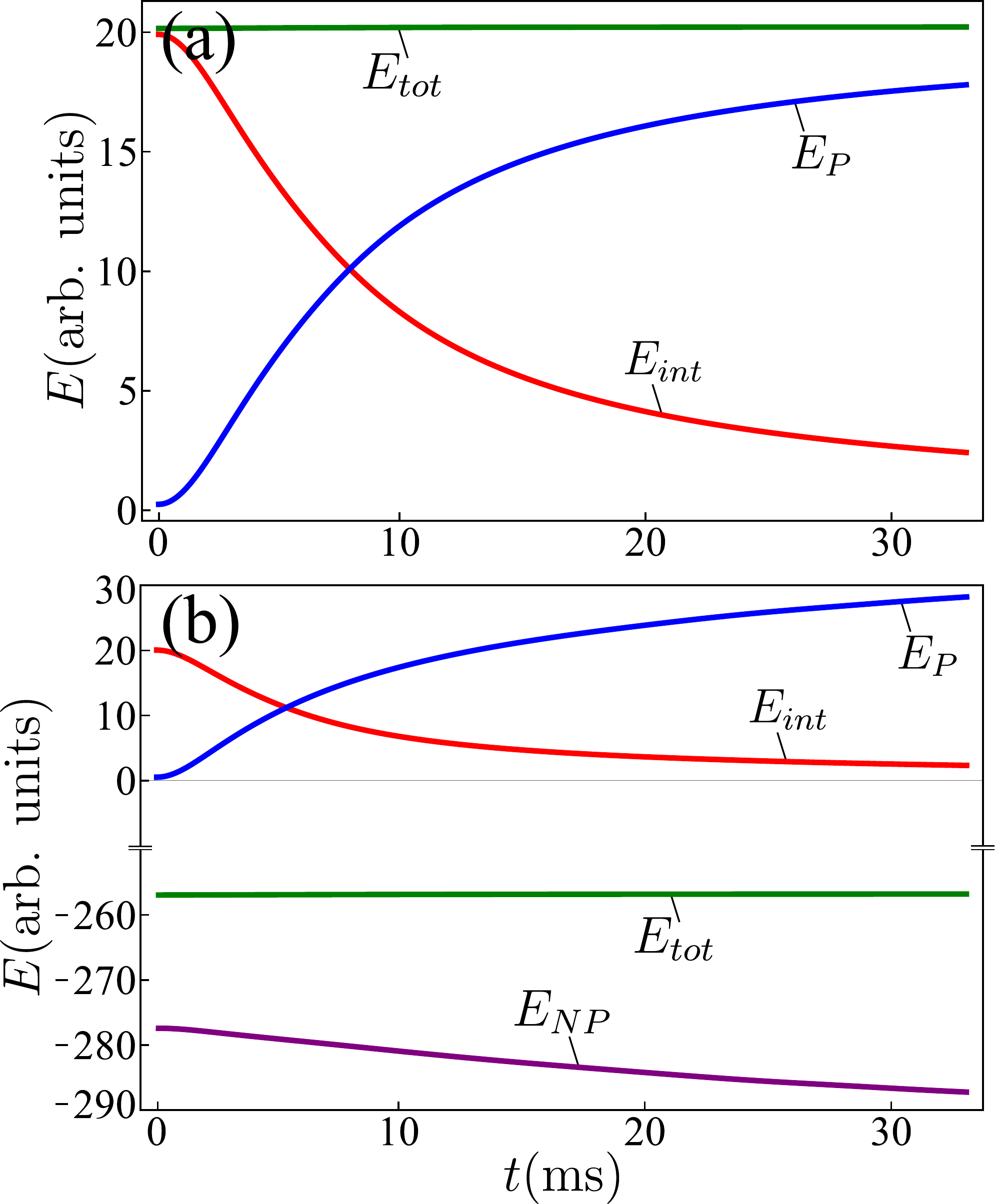}
  \caption{Energy conversions of the condensate (a) Parabolic dispersion case. At long times  the interaction energy $E_{\mathrm{int}}$ is entirely converted into kinetic energy $E_{\mathrm{P}}$. (b) SOCBEC dispersion case corresponding to  Fig.~2(b) of the main text. The interaction energy $E_{\mathrm{int}}$ is converted  into both the parabolic $E_{\mathrm{P}}$ and non-parabolic $E_{\mathrm{NP}}$ parts of the kinetic energy.}
  \label{fig:s_2}
\end{figure}
In Fig.~\ref{fig:s_2}(a) we illustrate the conversion of the interaction
energy into kinetic energy during the expansion of a single-component BEC described by a
 GPE with a parabolic dispersion. At long times the
interaction energy is entirely converted into kinetic energy, and the condensate behaves like a noninteracting Schr{\"o}dinger wave packet. The
case corresponding to the SOCBEC dispersion described in Fig.~2(c) of the main text is
presented in Fig.~\ref{fig:s_2}(b). The interaction energy is still
converted to kinetic energy, but this time with a significant contribution from the non-parabolic part of the dispersion.
\section{Analysis of experiment by Khamehchi \emph{et al.}}
\begin{figure}[h!]
  \includegraphics[width=\linewidth]{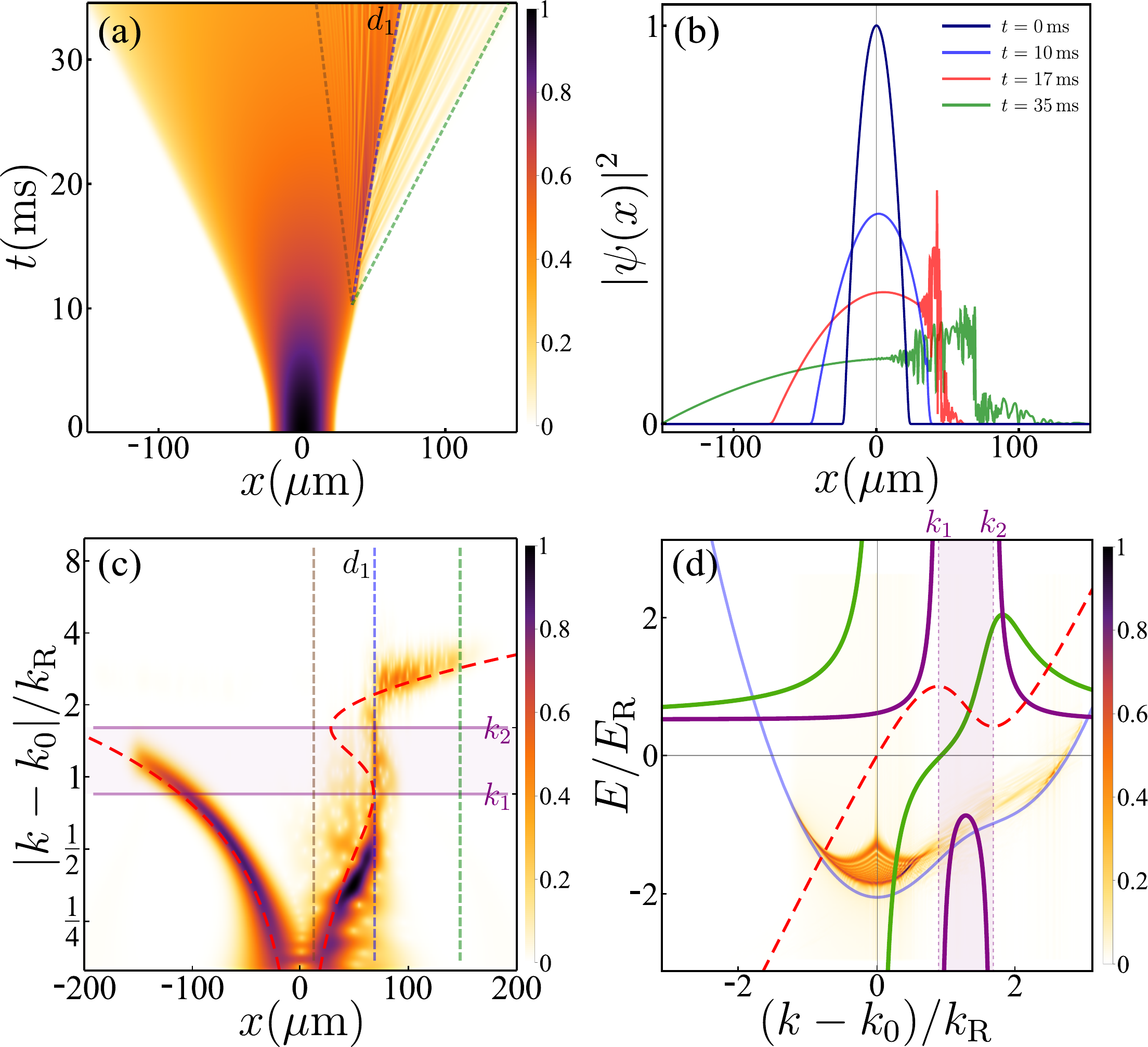}
  \caption{
Analysis of the dynamics of SOCBEC expansion as studied by Khamehchi \emph{et al.}~[18]. $\Omega=2.5$, $\delta=1.36$, $k_\mathrm{R}=1$, $\omega=2\pi\times \unit{26}{\hertz}$, Thomas-Fermi radius 23 $\mu$m. (a) Spacetime dynamics of the condensate density following release from the trap.  The brown and green dashed lines indicate the boundaries of self-interference, while the blue dashed line separates two distinct regions.  (b) Condensate density at $t=\{0, 10, 17, 35 \}$ ms after release.  (c) Wavelet Transform of the condensate wave function at $t=35$ ms.  The boundaries of self interference are indicated using the same line styles as in (a).  (d)  Blue line: Dispersion relation of the lower branch $E_{-}$.  Green line: effective mass $m_1$.  Purple line: effective mass $m_2$. Red dashed line: group velocity  $v$.  Overlaid is the condensate spectral density $|\psi(k,E)|^2$. An animation of this figure is provided as Supplementary Video 3.}
  \label{fig:s_3}
\end{figure}
In the main text of the paper by Khamehchi \emph{et al.}~[18], the authors present numerical results from three-dimensional simulations of the Gross-Pitaevskii equation for comparison with their experimental observations.  In addition, in the supplemental material they show that  one-dimensional simulations of a BEC with the same Thomas-Fermi radius exhibit very similar features~[18].  In this section we perform a wavelet analysis of the equivalent one-dimensional simulations to demonstrate that the root cause of their observations is the SIP phenomenon.

Our simulations begin with the ground state of a SOCBEC with $\Omega=2.5$, $\delta=1.36$, $k_\mathrm{R}=1$ in a $\omega=2\pi\times \unit{26}{\hertz}$ trap with a Thomas-Fermi radius of 23 $\mu$m, corresponding to approximately $10^5$ atoms. The spacetime evolution of the BEC density following release from the trap is shown in Fig.~\ref{fig:s_3}(a), with slices showing the condensate density at selected times in Fig.~\ref{fig:s_3}(b). The WT of the wave function taken at $t=t_{\textrm{max}}$ is shown in Fig.~\ref{fig:s_3}(c). The far-field of the wave function $\psi(x,t)$ can be computed by performing the Fourier Transform on both time and spatial coordinates, yielding the spectral density $|\psi(k,E)|^2$ and is shown in Fig.~\ref{fig:s_3}(d) overlaid with the dispersion relation for the system. This observable is
easily measured experimentally for polariton systems  where the photons leaking out of the cavity allow a continuous monitoring of the phase and density of the field. However such a measurement is challenging in atomic systems as the measurement of the density and phase of the condensate is generally destructive.  Nonetheless, there is no barrier to determining the condensate spectral density in simulations. 

From the spacetime density in Fig.~\ref{fig:s_3}(a) we can see two distinct regions of interference bounded on the outside by brown and green dashed lines, and separated in the middle by the blue dashed line labeled  $d_1$. The delimitation lines of the interference areas are also shown on the WT in  Fig.~\ref{fig:s_3}(c), where we can see that the ``leaking part'' previously identified by Khamehchi \textit{et al.}~[18] comes from the interference of $(k>k_2)$-components (when $m_2>0$ again) with lower $k-$components. We can also see in the comparison of the spectral density with the dispersion in Fig.~\ref{fig:s_3}(d) that the field follows the dispersion again at $k> k_2$. 

From this analysis we can see that the essential feature of the experiment of Khamehchi \textit{et al.}~[18] in comparison to the cases presented in the main text is a stronger nonlinearity, and a weaker initial confinement.  This means it takes longer for the generation of the higher-$k$ components, and  hence for the self-interference to occur.  The stronger nonlinearity leads to a more significant blue shift, and results in a more complex interference pattern combined with the the generation of grey solitons.

\section{Supplementary Videos}

Three supplementary videos are provided with this material.  The first two show the time-resolved SOCBEC dynamics in
both the linear and interacting regimes, as observed directly through
the wave packet motion in real space and the WTs.
Supplementary Video S1 is for the  SOC configuration of Khamehchi \emph{et al.}~[18] where only one mass
parameter is negative, and corresponds to the results presented in Fig.~2. Supplementary Video S2 is for the SOC configuration of the main text where both masses parameters are negative, corresponding to the results presented in Fig.~3.  Finally, Supplementary Video S3 shows the effect of the changing the strength of the nonlinearity on the SOCBEC dynamics corresponding to the experiment of Khamehchi \textit{et al.}~[18], and as presented here in Fig.~\ref{fig:s_3}.
The movie is parameterised by the Thomas-Fermi radius of the BEC, and begins from weakly interacting system with a small nonlinearity, and ends with the parameters of Fig.~\ref{fig:s_3}.

\begin{thebibliography}{28}
\expandafter\ifx\csname natexlab\endcsname\relax\def\natexlab#1{#1}\fi
\expandafter\ifx\csname bibnamefont\endcsname\relax
  \def\bibnamefont#1{#1}\fi
\expandafter\ifx\csname bibfnamefont\endcsname\relax
  \def\bibfnamefont#1{#1}\fi
\expandafter\ifx\csname citenamefont\endcsname\relax
  \def\citenamefont#1{#1}\fi
\expandafter\ifx\csname url\endcsname\relax
  \def\url#1{\texttt{#1}}\fi
\expandafter\ifx\csname urlprefix\endcsname\relax\def\urlprefix{URL }\fi
\providecommand{\bibinfo}[2]{#2}
\providecommand{\eprint}[2][]{\url{#2}}

\bibitem[{\citenamefont{Kittel}(2004)}]{kittel_book04a}
\bibinfo{author}{\bibfnamefont{C.}~\bibnamefont{Kittel}},
  \emph{\bibinfo{title}{Introduction to Solid State Physics}}
  (\bibinfo{publisher}{Wiley}, \bibinfo{year}{2004}), \bibinfo{edition}{8th}
  ed.

\bibitem[{\citenamefont{Larson et~al.}(2005)\citenamefont{Larson, Salo, and
  Stenholm}}]{larson05a}
\bibinfo{author}{\bibfnamefont{J.}~\bibnamefont{Larson}},
  \bibinfo{author}{\bibfnamefont{J.}~\bibnamefont{Salo}}, \bibnamefont{and}
  \bibinfo{author}{\bibfnamefont{S.}~\bibnamefont{Stenholm}},
  \bibinfo{journal}{Phys. Rev. A} \textbf{\bibinfo{volume}{72}},
  \bibinfo{pages}{013814} (\bibinfo{year}{2005}).

\bibitem[{\citenamefont{Egorov et~al.}(2009)\citenamefont{Egorov, Skryabin,
  Yulin, and Lederer}}]{egorov09a}
\bibinfo{author}{\bibfnamefont{O.~A.} \bibnamefont{Egorov}},
  \bibinfo{author}{\bibfnamefont{D.~V.} \bibnamefont{Skryabin}},
  \bibinfo{author}{\bibfnamefont{A.~V.} \bibnamefont{Yulin}}, \bibnamefont{and}
  \bibinfo{author}{\bibfnamefont{F.}~\bibnamefont{Lederer}},
  \bibinfo{journal}{Phys. Rev. Lett.} \textbf{\bibinfo{volume}{102}},
  \bibinfo{pages}{153904} (\bibinfo{year}{2009}).

\bibitem[{\citenamefont{Colas and Laussy}(2016)}]{colas16a}
\bibinfo{author}{\bibfnamefont{D.}~\bibnamefont{Colas}} \bibnamefont{and}
  \bibinfo{author}{\bibfnamefont{F.~P.} \bibnamefont{Laussy}},
  \bibinfo{journal}{Phys. Rev. Lett.} \textbf{\bibinfo{volume}{116}},
  \bibinfo{pages}{026401} (\bibinfo{year}{2016}).

\bibitem[{\citenamefont{Kavokin et~al.}(2017)\citenamefont{Kavokin, Baumberg,
  Malpuech, and Laussy}}]{kavokin_book17a}
\bibinfo{author}{\bibfnamefont{A.}~\bibnamefont{Kavokin}},
  \bibinfo{author}{\bibfnamefont{J.~J.} \bibnamefont{Baumberg}},
  \bibinfo{author}{\bibfnamefont{G.}~\bibnamefont{Malpuech}}, \bibnamefont{and}
  \bibinfo{author}{\bibfnamefont{F.~P.} \bibnamefont{Laussy}},
  \emph{\bibinfo{title}{Microcavities}} (\bibinfo{publisher}{Oxford University
  Press}, \bibinfo{year}{2017}), \bibinfo{edition}{2nd} ed.

\bibitem[{\citenamefont{Savvidis et~al.}(2000)\citenamefont{Savvidis, Baumberg,
  Stevenson, Skolnick, Whittaker, and Roberts}}]{savvidis00a}
\bibinfo{author}{\bibfnamefont{P.~G.} \bibnamefont{Savvidis}},
  \bibinfo{author}{\bibfnamefont{J.~J.} \bibnamefont{Baumberg}},
  \bibinfo{author}{\bibfnamefont{R.~M.} \bibnamefont{Stevenson}},
  \bibinfo{author}{\bibfnamefont{M.~S.} \bibnamefont{Skolnick}},
  \bibinfo{author}{\bibfnamefont{D.~M.} \bibnamefont{Whittaker}},
  \bibnamefont{and} \bibinfo{author}{\bibfnamefont{J.~S.}
  \bibnamefont{Roberts}}, \bibinfo{journal}{Phys. Rev. Lett.}
  \textbf{\bibinfo{volume}{84}}, \bibinfo{pages}{1547} (\bibinfo{year}{2000}).

\bibitem[{\citenamefont{Amo et~al.}(2009)\citenamefont{Amo, Sanvitto, Laussy,
  Ballarini, del Valle, Martin, Lema\^itre, Bloch, Krizhanovskii, Skolnick
  et~al.}}]{amo09a}
\bibinfo{author}{\bibfnamefont{A.}~\bibnamefont{Amo}},
  \bibinfo{author}{\bibfnamefont{D.}~\bibnamefont{Sanvitto}},
  \bibinfo{author}{\bibfnamefont{F.~P.} \bibnamefont{Laussy}},
  \bibinfo{author}{\bibfnamefont{D.}~\bibnamefont{Ballarini}},
  \bibinfo{author}{\bibfnamefont{E.}~\bibnamefont{del Valle}},
  \bibinfo{author}{\bibfnamefont{M.~D.} \bibnamefont{Martin}},
  \bibinfo{author}{\bibfnamefont{A.}~\bibnamefont{Lema\^itre}},
  \bibinfo{author}{\bibfnamefont{J.}~\bibnamefont{Bloch}},
  \bibinfo{author}{\bibfnamefont{D.~N.} \bibnamefont{Krizhanovskii}},
  \bibinfo{author}{\bibfnamefont{M.~S.} \bibnamefont{Skolnick}},
  \bibnamefont{et~al.}, \bibinfo{journal}{Nature}
  \textbf{\bibinfo{volume}{457}}, \bibinfo{pages}{291} (\bibinfo{year}{2009}).

\bibitem[{\citenamefont{Tosi et~al.}(2011)\citenamefont{Tosi, Marchetti,
  Sanvitto, Ant\'on, Szyma\'nska, Berceanu, Tejedor, Marrucci, Lema\^{\i}tre,
  Bloch et~al.}}]{tosi11a}
\bibinfo{author}{\bibfnamefont{G.}~\bibnamefont{Tosi}},
  \bibinfo{author}{\bibfnamefont{F.~M.} \bibnamefont{Marchetti}},
  \bibinfo{author}{\bibfnamefont{D.}~\bibnamefont{Sanvitto}},
  \bibinfo{author}{\bibfnamefont{C.}~\bibnamefont{Ant\'on}},
  \bibinfo{author}{\bibfnamefont{M.~H.} \bibnamefont{Szyma\'nska}},
  \bibinfo{author}{\bibfnamefont{A.}~\bibnamefont{Berceanu}},
  \bibinfo{author}{\bibfnamefont{C.}~\bibnamefont{Tejedor}},
  \bibinfo{author}{\bibfnamefont{L.}~\bibnamefont{Marrucci}},
  \bibinfo{author}{\bibfnamefont{A.}~\bibnamefont{Lema\^{\i}tre}},
  \bibinfo{author}{\bibfnamefont{J.}~\bibnamefont{Bloch}},
  \bibnamefont{et~al.}, \bibinfo{journal}{Phys. Rev. Lett.}
  \textbf{\bibinfo{volume}{107}}, \bibinfo{pages}{036401}
  (\bibinfo{year}{2011}).

\bibitem[{\citenamefont{Ballarini et~al.}(2017)\citenamefont{Ballarini, Caputo,
  Mu{\~ n}oz, Giorgi, Dominici, Szyma\'nska, West, Pfeiffer, Gigli, Laussy
  et~al.}}]{ballarini17a}
\bibinfo{author}{\bibfnamefont{D.}~\bibnamefont{Ballarini}},
  \bibinfo{author}{\bibfnamefont{D.}~\bibnamefont{Caputo}},
  \bibinfo{author}{\bibfnamefont{C.~S.} \bibnamefont{Mu{\~ n}oz}},
  \bibinfo{author}{\bibfnamefont{M.~D.} \bibnamefont{Giorgi}},
  \bibinfo{author}{\bibfnamefont{L.}~\bibnamefont{Dominici}},
  \bibinfo{author}{\bibfnamefont{M.~H.} \bibnamefont{Szyma\'nska}},
  \bibinfo{author}{\bibfnamefont{K.}~\bibnamefont{West}},
  \bibinfo{author}{\bibfnamefont{L.~N.} \bibnamefont{Pfeiffer}},
  \bibinfo{author}{\bibfnamefont{G.}~\bibnamefont{Gigli}},
  \bibinfo{author}{\bibfnamefont{F.~P.} \bibnamefont{Laussy}},
  \bibnamefont{et~al.}, \bibinfo{journal}{Phys. Rev. Lett.}
  \textbf{\bibinfo{volume}{118}}, \bibinfo{pages}{215301}
  (\bibinfo{year}{2017}).

\bibitem[{\citenamefont{Gianfrate et~al.}(2018)\citenamefont{Gianfrate,
  Dominici, Voronych, Matuszewski, Stobi\`{n}ska, Ballarini, Giorgi, Gigli, and
  Sanvitto}}]{gianfrate18}
\bibinfo{author}{\bibfnamefont{A.}~\bibnamefont{Gianfrate}},
  \bibinfo{author}{\bibfnamefont{L.}~\bibnamefont{Dominici}},
  \bibinfo{author}{\bibfnamefont{O.}~\bibnamefont{Voronych}},
  \bibinfo{author}{\bibfnamefont{M.}~\bibnamefont{Matuszewski}},
  \bibinfo{author}{\bibfnamefont{M.}~\bibnamefont{Stobi\`{n}ska}},
  \bibinfo{author}{\bibfnamefont{D.}~\bibnamefont{Ballarini}},
  \bibinfo{author}{\bibfnamefont{M.~D.} \bibnamefont{Giorgi}},
  \bibinfo{author}{\bibfnamefont{G.}~\bibnamefont{Gigli}}, \bibnamefont{and}
  \bibinfo{author}{\bibfnamefont{D.}~\bibnamefont{Sanvitto}},
  \bibinfo{journal}{Light: Sci. \& App.} \textbf{\bibinfo{volume}{7}},
  \bibinfo{pages}{17119} (\bibinfo{year}{2018}).

\bibitem[{\citenamefont{Eiermann et~al.}(2003)\citenamefont{Eiermann,
  Treutlein, Anker, Albiez, Taglieber, Marzlin, and Oberthaler}}]{eiermann03a}
\bibinfo{author}{\bibfnamefont{B.}~\bibnamefont{Eiermann}},
  \bibinfo{author}{\bibfnamefont{P.}~\bibnamefont{Treutlein}},
  \bibinfo{author}{\bibfnamefont{T.}~\bibnamefont{Anker}},
  \bibinfo{author}{\bibfnamefont{M.}~\bibnamefont{Albiez}},
  \bibinfo{author}{\bibfnamefont{M.}~\bibnamefont{Taglieber}},
  \bibinfo{author}{\bibfnamefont{K.-P.} \bibnamefont{Marzlin}},
  \bibnamefont{and} \bibinfo{author}{\bibfnamefont{M.~K.}
  \bibnamefont{Oberthaler}}, \bibinfo{journal}{Phys. Rev. Lett.}
  \textbf{\bibinfo{volume}{91}}, \bibinfo{pages}{060402}
  (\bibinfo{year}{2003}).

\bibitem[{\citenamefont{Morsch and Oberthaler}(2006)}]{morsch06a}
\bibinfo{author}{\bibfnamefont{O.}~\bibnamefont{Morsch}} \bibnamefont{and}
  \bibinfo{author}{\bibfnamefont{M.}~\bibnamefont{Oberthaler}},
  \bibinfo{journal}{Rev. Mod. Phys.} \textbf{\bibinfo{volume}{78}},
  \bibinfo{pages}{179} (\bibinfo{year}{2006}).

\bibitem[{\citenamefont{Eiermann et~al.}(2004)\citenamefont{Eiermann, Anker,
  Albiez, Taglieber, Treutlein, Marzlin, and Oberthaler}}]{eiermann04a}
\bibinfo{author}{\bibfnamefont{B.}~\bibnamefont{Eiermann}},
  \bibinfo{author}{\bibfnamefont{T.}~\bibnamefont{Anker}},
  \bibinfo{author}{\bibfnamefont{M.}~\bibnamefont{Albiez}},
  \bibinfo{author}{\bibfnamefont{M.}~\bibnamefont{Taglieber}},
  \bibinfo{author}{\bibfnamefont{P.}~\bibnamefont{Treutlein}},
  \bibinfo{author}{\bibfnamefont{K.-P.} \bibnamefont{Marzlin}},
  \bibnamefont{and} \bibinfo{author}{\bibfnamefont{M.~K.}
  \bibnamefont{Oberthaler}}, \bibinfo{journal}{Phys. Rev. Lett.}
  \textbf{\bibinfo{volume}{92}}, \bibinfo{pages}{230401}
  (\bibinfo{year}{2004}).

\bibitem[{\citenamefont{Sich et~al.}(2012)\citenamefont{Sich, Krizhanovskii,
  Skolnick, Gorbach, Hartley, Skryabin, Cerda-M\'endez, Biermann, Hey, and
  Santos}}]{sich12a}
\bibinfo{author}{\bibfnamefont{M.}~\bibnamefont{Sich}},
  \bibinfo{author}{\bibfnamefont{D.}~\bibnamefont{Krizhanovskii}},
  \bibinfo{author}{\bibfnamefont{M.}~\bibnamefont{Skolnick}},
  \bibinfo{author}{\bibfnamefont{A.}~\bibnamefont{Gorbach}},
  \bibinfo{author}{\bibfnamefont{R.}~\bibnamefont{Hartley}},
  \bibinfo{author}{\bibfnamefont{D.~V.} \bibnamefont{Skryabin}},
  \bibinfo{author}{\bibfnamefont{E.~A.} \bibnamefont{Cerda-M\'endez}},
  \bibinfo{author}{\bibfnamefont{K.}~\bibnamefont{Biermann}},
  \bibinfo{author}{\bibfnamefont{R.}~\bibnamefont{Hey}}, \bibnamefont{and}
  \bibinfo{author}{\bibfnamefont{P.}~\bibnamefont{Santos}},
  \bibinfo{journal}{Nat. Photon.} \textbf{\bibinfo{volume}{6}},
  \bibinfo{pages}{50} (\bibinfo{year}{2012}).

\bibitem[{\citenamefont{Lin et~al.}(2011)\citenamefont{Lin, Jimenez-Garcia, and
  Spielman}}]{linYJ11a}
\bibinfo{author}{\bibfnamefont{Y.~J.} \bibnamefont{Lin}},
  \bibinfo{author}{\bibfnamefont{K.}~\bibnamefont{Jimenez-Garcia}},
  \bibnamefont{and} \bibinfo{author}{\bibfnamefont{I.~B.}
  \bibnamefont{Spielman}}, \bibinfo{journal}{Nature}
  \textbf{\bibinfo{volume}{471}}, \bibinfo{pages}{83} (\bibinfo{year}{2011}).

\bibitem[{\citenamefont{Zhang et~al.}(2012)\citenamefont{Zhang, Mao, and
  Zhang}}]{zhang12b}
\bibinfo{author}{\bibfnamefont{Y.}~\bibnamefont{Zhang}},
  \bibinfo{author}{\bibfnamefont{L.}~\bibnamefont{Mao}}, \bibnamefont{and}
  \bibinfo{author}{\bibfnamefont{C.}~\bibnamefont{Zhang}},
  \bibinfo{journal}{Phys. Rev. Lett.} \textbf{\bibinfo{volume}{108}},
  \bibinfo{pages}{035302} (\bibinfo{year}{2012}).

\bibitem[{\citenamefont{Hamner et~al.}(2015)\citenamefont{Hamner, Zhang,
  Khamehchi, Davis, and Engels}}]{hamner15a}
\bibinfo{author}{\bibfnamefont{C.}~\bibnamefont{Hamner}},
  \bibinfo{author}{\bibfnamefont{Y.}~\bibnamefont{Zhang}},
  \bibinfo{author}{\bibfnamefont{M.~A.} \bibnamefont{Khamehchi}},
  \bibinfo{author}{\bibfnamefont{M.~J.} \bibnamefont{Davis}}, \bibnamefont{and}
  \bibinfo{author}{\bibfnamefont{P.}~\bibnamefont{Engels}},
  \bibinfo{journal}{Phys. Rev. Lett.} \textbf{\bibinfo{volume}{114}},
  \bibinfo{pages}{070401} (\bibinfo{year}{2015}).

\bibitem[{\citenamefont{Khamehchi et~al.}(2017)\citenamefont{Khamehchi,
  Hossain, Mossman, Zhang, Busch, Forbes, and Engels}}]{khamehchi17a}
\bibinfo{author}{\bibfnamefont{M.~A.} \bibnamefont{Khamehchi}},
  \bibinfo{author}{\bibfnamefont{K.}~\bibnamefont{Hossain}},
  \bibinfo{author}{\bibfnamefont{M.~E.} \bibnamefont{Mossman}},
  \bibinfo{author}{\bibfnamefont{Y.}~\bibnamefont{Zhang}},
  \bibinfo{author}{\bibfnamefont{T.}~\bibnamefont{Busch}},
  \bibinfo{author}{\bibfnamefont{M.~M.} \bibnamefont{Forbes}},
  \bibnamefont{and} \bibinfo{author}{\bibfnamefont{P.}~\bibnamefont{Engels}},
  \bibinfo{journal}{Phys. Rev. Lett.} \textbf{\bibinfo{volume}{118}},
  \bibinfo{pages}{155301} (\bibinfo{year}{2017}).

\bibitem[{Sup({\natexlab{a}})}]{SuppMat}
\bibinfo{note}{See Supplemental Material for a detailed analysis on the
  dispersion properties, details on the Wavelet Transform applied to wave
  packets, a further description of the expansion of the BEC that broadens the
  momentum space wave packet, the comparison with the experimental case, and a
  description of the supplementary video.}

\bibitem[{Sup({\natexlab{b}})}]{SuppMat2}
\bibinfo{note}{\href{http://demonstrations.wolfram.com/DispersionPropertiesOfASpinOrbitCoupledBoseEinsteinCondensat/}{http://demonstrations.wolfram.com/
  DispersionPropertiesOfASpinOrbitCoupledBoseEinsteinCondensat/}}.

\bibitem[{foo({\natexlab{a}})}]{footnote1}
\bibinfo{note}{We note that Ref.~[18] showed that simulations of the 1D and 3D
  Gross-Pitaevskii equations gave quantitatively similar results.}

\bibitem[{foo({\natexlab{b}})}]{footnote2}
\bibinfo{note}{For $^{87}$Rb this width would require a harmonic trap of
  frequency 1860 Hz for $g=0$.}

\bibitem[{\citenamefont{Debnath and Shah}(2015)}]{debnath_book15a}
\bibinfo{author}{\bibfnamefont{L.}~\bibnamefont{Debnath}} \bibnamefont{and}
  \bibinfo{author}{\bibfnamefont{F.~A.} \bibnamefont{Shah}},
  \emph{\bibinfo{title}{Wavelet Transforms and Their Applications}}
  (\bibinfo{publisher}{Birkh{\u a}user}, \bibinfo{year}{2015}),
  \bibinfo{edition}{2nd} ed.

\bibitem[{\citenamefont{Baker et~al.}(2012)\citenamefont{Baker, Jordan, and
  Norris}}]{baker12a}
\bibinfo{author}{\bibfnamefont{C.~H.} \bibnamefont{Baker}},
  \bibinfo{author}{\bibfnamefont{D.~A.} \bibnamefont{Jordan}},
  \bibnamefont{and} \bibinfo{author}{\bibfnamefont{P.~M.}
  \bibnamefont{Norris}}, \bibinfo{journal}{Phys. Rev. B}
  \textbf{\bibinfo{volume}{86}}, \bibinfo{pages}{104306}
  (\bibinfo{year}{2012}).

\bibitem[{\citenamefont{M\'ark}(1997)}]{mark97a}
\bibinfo{author}{\bibfnamefont{G.~I.} \bibnamefont{M\'ark}},
  \bibinfo{journal}{Eur. Phys. J. B} \textbf{\bibinfo{volume}{18}},
  \bibinfo{pages}{247} (\bibinfo{year}{1997}).

\bibitem[{\citenamefont{Anker et~al.}(2005)\citenamefont{Anker, Albiez, Gati,
  Hunsmann, Eiermann, Trombettoni, and Oberthaler}}]{anker05a}
\bibinfo{author}{\bibfnamefont{T.}~\bibnamefont{Anker}},
  \bibinfo{author}{\bibfnamefont{M.}~\bibnamefont{Albiez}},
  \bibinfo{author}{\bibfnamefont{R.}~\bibnamefont{Gati}},
  \bibinfo{author}{\bibfnamefont{S.}~\bibnamefont{Hunsmann}},
  \bibinfo{author}{\bibfnamefont{B.}~\bibnamefont{Eiermann}},
  \bibinfo{author}{\bibfnamefont{A.}~\bibnamefont{Trombettoni}},
  \bibnamefont{and} \bibinfo{author}{\bibfnamefont{M.~K.}
  \bibnamefont{Oberthaler}}, \bibinfo{journal}{Phys. Rev. Lett.}
  \textbf{\bibinfo{volume}{94}}, \bibinfo{pages}{020403}
  (\bibinfo{year}{2005}).

\bibitem[{\citenamefont{Alexander et~al.}(2006)\citenamefont{Alexander,
  Ostrovskaya, and Kivshar}}]{alexander06a}
\bibinfo{author}{\bibfnamefont{T.~J.} \bibnamefont{Alexander}},
  \bibinfo{author}{\bibfnamefont{E.~A.} \bibnamefont{Ostrovskaya}},
  \bibnamefont{and} \bibinfo{author}{\bibfnamefont{Y.~S.}
  \bibnamefont{Kivshar}}, \bibinfo{journal}{Phys. Rev. Lett.}
  \textbf{\bibinfo{volume}{96}}, \bibinfo{pages}{040401}
  (\bibinfo{year}{2006}).

\bibitem[{\citenamefont{Xue et~al.}(2008)\citenamefont{Xue, Zhang, and
  Liu}}]{xue08a}
\bibinfo{author}{\bibfnamefont{J.~K.} \bibnamefont{Xue}},
  \bibinfo{author}{\bibfnamefont{A.~X.} \bibnamefont{Zhang}}, \bibnamefont{and}
  \bibinfo{author}{\bibfnamefont{J.}~\bibnamefont{Liu}},
  \bibinfo{journal}{Phys. Rev. A} \textbf{\bibinfo{volume}{77}},
  \bibinfo{pages}{013602} (\bibinfo{year}{2008}).

\end{thebibliography}
\end{document}